\documentclass[aip,jap,reprint,graphicx]{revtex4-1} % for checking your page length
\usepackage{graphicx,epsf,epsfig}% Include figure files
\usepackage{dcolumn}% Align table columns on decimal point
\usepackage{bm}% bold math
\usepackage{amsmath, amsthm, amssymb}
\usepackage{color}
\usepackage{ulem}

\usepackage{epstopdf}
\usepackage{float}
\usepackage{longtable}
\usepackage{multirow}
\begin{document}

\title{FORC+: A method for separating reversible from irreversible behavior using first order reversal curves}
%\thanks{}
\author{P. B. Visscher}
\affiliation{Center for Materials for Information Technology, U. of Alabama, Tuscaloosa, AL 35401 USA}
\affiliation{Department of Physics and Astronomy, Univ. of Alabama, Tuscaloosa, AL 35401, USA}

%\widetext
\begin{abstract}
First Order Reversal Curves (FORCs) have been used for a number of years for the extraction of information from magnetization measurements.  The results are most unambiguous for irreversible processes -- for a collection of Preisach hysterons, one gets a "FORC distribution" $\rho(H_{down},H_{up})$, the number of hysterons with given downward \& upward reversal fields.  There have been many proposals for dealing with reversible behavior, usually involving inserting it somehow into the irreversible FORC distribution.  Here we will try to do the opposite, to separate them into another function which we will call the (reversible) "saturation field distribution", which is identically zero for a completely irreversible system of hysterons, while the irreversible FORC distribution is identically zero for a reversible system.  Thus in a system with both purely reversible and purely irreversible components, such as single-domain Stoner-Wohlfarth particles with hard or easy axis along the field, this approach cleanly separates them.  For more complicated systems, as with conventional FORC distributions, it at least provides a "signature" making it possible to identify microscopic models that might give a particular pair of irreversible and reversible distributions.
\end{abstract}

\maketitle
% keywords: FORC, reversible

\section{Introduction}
\begin{figure}[htb]
\begin{center}
\includegraphics[width=2.5in]{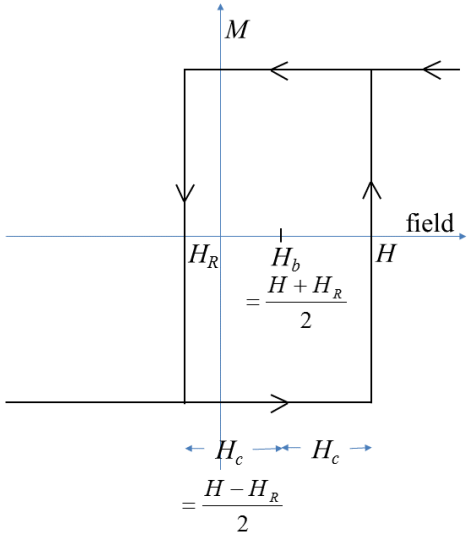}
\caption{\label{hyst} MH loop of a single Preisach hysteron, showing down-switching field $H_R$ and up-switching field $H$, and defining the bias field $H_b$ and the coercivity $H_c$.}
\end{center}
%\label{figure:MI} this causes references to give section number, not figure number!!
%\vspace{-3mm}
\end{figure}

\begin{figure}[htb]
\begin{center}
\includegraphics[width=3in]{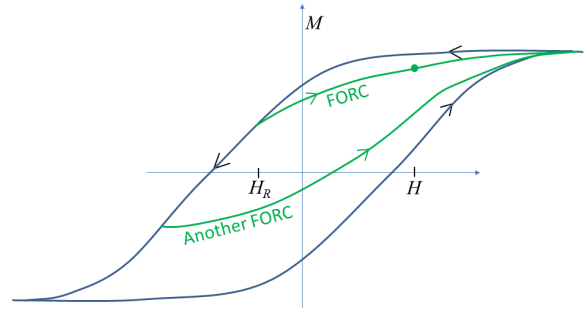}
\caption{\label{FORC} A major hysteresis loop with two FORC curves, with a dot showing the point where $M(H_R, H)$ is defined.}
\end{center}
\end{figure}

The FORC method\cite{Pike2003,PikeNi2005,Stancu2013,RobertsReview2014} was originally designed for completely irreversible systems, modeled as a collection of Preisach hysterons, each of which has a rectangular hysteresis loop (Fig. \ref{hyst}).  As we lower the field from a large positive saturating value, it switches down at a field usually denoted by $H_R$, (the subscript R stands for "reversal", for reasons that will become apparent later) and as we increase the field it switches back up at a field $H > H_R$. The Preisach distribution is the density of these hysterons in the $H - H_R$ plane (Fig. \ref{prei1}).
The fundamental result behind the FORC idea is that this Preisach distribution can be obtained by measuring "first order reversal curves" -- that is, by saturating the sample in the positive direction, decreasing the field to a reversal field $H_R$ (see Fig. \ref{FORC}), then reversing dH/dt from negative to positive and measuring the magnetization $M(H_R,H)$ as the field increases again past each value $H$.

The distribution of hysterons is then given by
\begin{equation}\label{dist}
\rho(H_R,H)=-\frac{1}{2} \frac{\partial^2\mathbf{M}(H_R,H)}{\partial H_R \partial H}
\end{equation}
In Sec. \ref{deriv} we will give a derivation of this result, using a discrete formulation.  We will show that this distribution fails to include reversible effects, and show that the information not included in $\rho(H_R,H)$ can be thought of as the first derivative $\partial\mathbf{M}(H_R,H) / \partial H$ evaluated at $H=H_R$, which vanishes in an irreversible system of Preisach hysterons with nonzero coercivity.  For the case of purely reversible hard axis particles characterized by saturation fields, we show that a derivative of this quantity with respect to $H_R$ can be interpreted as a distribution of saturation fields.  Thus we can extract the distribution of both irreversible and reversible particles from the FORC data.

\section{Discrete derivation of FORC distribution} \label{deriv}
\begin{figure}[htb]
\begin{center}
\includegraphics[width=3in]{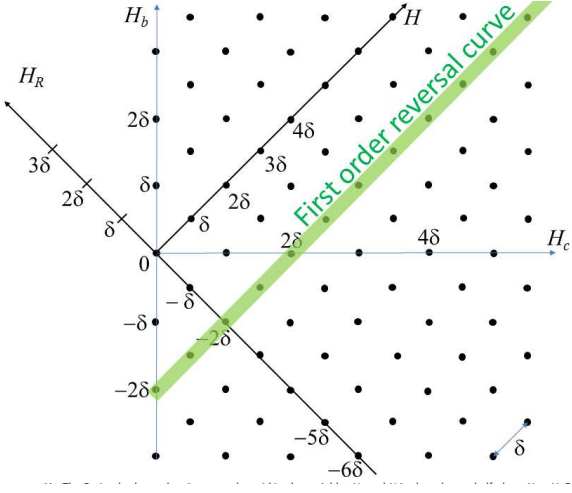}
\caption{\label{prei1} The Preisach plane, showing a regular grid in the variables $H_R$ and $H$ in the relevant half-plane $H_R \leq H$.  It is conventional to draw the $H$ and $H_R$ axes diagonally so the coercivity and bias field axes (defined in Fig. \ref{hyst}) can be horizontal and vertical.  The points inside the green stripe make up a single FORC curve at $H_R=-2\delta$. }
\end{center}
%\label{figure:MI} this causes references to give section number, not figure number!!
%\vspace{-3mm}
\end{figure}

We will begin by giving a simple derivation of Eq. \ref{dist}, relating the distribution of hysterons to a mixed partial derivative of the FORC function.  Although this formula is given in every paper on FORC, it is surprisingly hard to find a derivation.   Rather than give a derivation in the continuum limit, we will derive a discrete analog on a grid with a finite field spacing $\delta$ (Fig. \ref{prei1}), which becomes Eq. \ref{dist} in the limit $\delta \rightarrow 0$ but is easier to visualize.
Fig. \ref{prei1} shows the points in the $H - H_R$ plane at which the FORC function $\mathbf{M}(H_R,H)$ is measured.

\begin{figure}[htb]
\begin{center}
\includegraphics[width=3in]{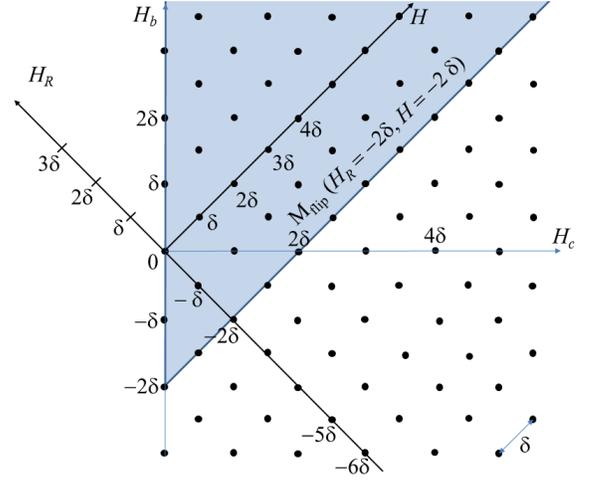}
\caption{\label{HR} Preisach plane after reducing the field to $H_R = -2\delta$.  Hysterons in blue-shaded region have been switched down. Total magnetic moment is now $M(-2\delta,-2\delta)$.}
\end{center}
%\label{figure:MI} this causes references to give section number, not figure number!!
%\vspace{-3mm}
\end{figure}

When we begin a FORC curve by reducing the field to $H_R$, for example $H_R=-2\delta$ as shown in Fig. \ref{prei1}, we flip downward all the hysterons having $H_R>-2\delta$, i. e., those in the blue area of Fig. \ref{HR}, whose total saturation moment we will denote by $M_{flip}(H_R,H_R)$.

It is related to the total remaining moment $M(H_R,H_R)=M_s - 2 M_{flip}(H_R,H_R)$ (the factor of $-2$ is because flipping an object with saturation moment $M_s$ changes the total moment by $-2M_s$).
\begin{figure}[htb]
\begin{center}
\includegraphics[width=3in]{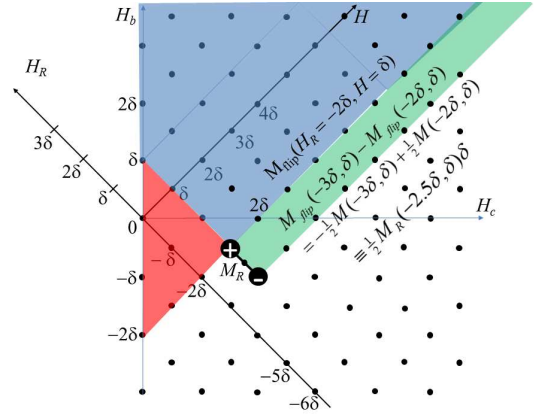}
\caption{\label{Strip} Preisach plane after raising the field from $H_R=-2\delta$ to $H=\delta$ -- the hysterons in the pink triangle have switched back up, leaving moment (shaded in blue) $M(-2\delta,\delta)$.  The green region is the additional area that would be flipped if we used $H_R=-3\delta$ instead. }
\end{center}
%\label{figure:MI} this causes references to give section number, not figure number!!
%\vspace{-3mm}
\end{figure}

\begin{figure}[htb]
\begin{center}
\includegraphics[width=3in]{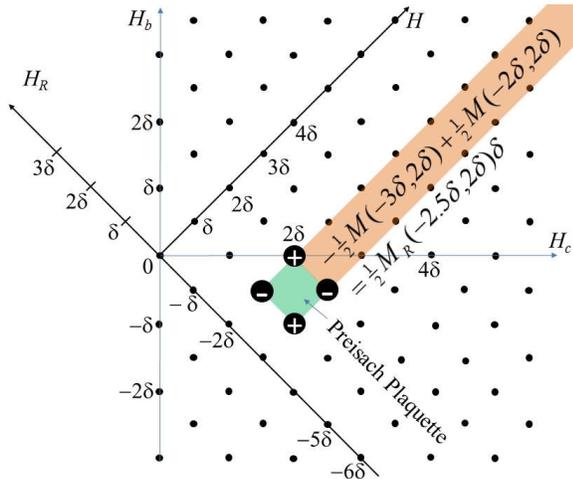}
\caption{\label{Plaq} Graphical demonstration that the saturation moment in the green Preisach plaquette is $\frac{1}{2} [ M(H_R-\frac{1}{2}\delta, H-\frac{1}{2}\delta) - M(H_R+\frac{1}{2}\delta, H-\frac{1}{2}\delta)+M(H_R+\frac{1}{2}\delta, H+\frac{1}{2}\delta)-M(H_R-\frac{1}{2}\delta, H+\frac{1}{2}\delta)  ]$ for $H_R^p=-2.5\delta, H^p=1.5\delta)$.  The signs on the four black dots indicate the signs of the four terms. }
\end{center}
%\label{figure:MI} this causes references to give section number, not figure number!!
%\vspace{-3mm}
\end{figure}

If we then increase the field to $H=\delta$, the hysterons in the pink triangle of Fig. \ref{Strip}, which have upward switching field $H<\delta$, switch back up, leaving only the hysterons in the blue area of Fig. \ref{Strip} flipped, whose total moment we denote by $M_{flip}(H_R,H)$, giving overall system moment
\begin{equation}\label{flip}
M(H_R,H)=M_s - 2 M_{flip}(H_R,H)
\end{equation}
If we now repeat this process with a smaller $H_R = -3\delta$, the additional hysterons in the green strip in Fig. \ref{Strip} will have flipped, with total moment $M_{flip}(-3\delta,\delta)$, so the moment in the green strip is $M_{flip}(-3\delta,\delta)-M_{flip}(-2\delta,\delta)$.  Expressing this difference in terms of the FORC function $M(H_R,H)$ by using Eq. \ref{flip}, the $M_s$ cancels and we get $\frac{1}{2} M(-3\delta,\delta) - \frac{1}{2} M(-2\delta,\delta)$, as indicated in Fig. \ref{Strip}, which can be expressed in terms of the discrete derivative $\partial M/\partial H_R$, which we denote by $M_R$ and define by
\begin{equation}\label{MR}
M_R(H_R,H)\equiv \frac{1}{\delta} [ M(H_R+\frac{1}{2}\delta, H) - M(H_R-\frac{1}{2}\delta, H) ]
\end{equation}
This definition is indicated pictorially in Fig. \ref{Strip} by a dumbbell labeled $M_R$ with + and - signs at the points where $M$ is to be added and subtracted.
If we repeat this process again with a larger $H=2\delta$, we will get the moment of the orange strip in Fig. \ref{Plaq}, which is $\frac{1}{2} M_R(-2.5\delta,2\delta)\delta$.
The difference, the saturation moment of the hysterons in the green square ("Preisach plaquette") in Fig. \ref{Plaq}, is then $-\delta^2$ times a second derivative
\begin{equation}\label{MRH}
M_{RH}(H_R,H)\equiv \frac{1}{\delta} [ M_R(H_R, H+\frac{1}{2}\delta) - M_R(H_R, H-\frac{1}{2}\delta) ]
\end{equation}
We define a (Preisach) density of hysterons such that the total saturation moment in a plaquette centered at $(H_R^p,H^p)$ is $\rho(H_R^p,H^p) \delta^2$.

We include the factor $\delta^2$ so that $\rho$ has units of magnetic moment/(field)$^2$, and $\rho$ is independent of $\delta$ in the limit $\delta \rightarrow 0$.  Then we have
\begin{equation}\label{rho}
\rho(H_R^p,H^p) = -\frac{1}{2} M_{RH}(H_R^p,H^p)
\end{equation}
which becomes the continuum Equation (\ref{dist}) in the limit $\delta \rightarrow 0$.

\section{Reversible behavior}
\label{next}
\begin{figure}[htb]
\begin{center}
\includegraphics[width=2.5in]{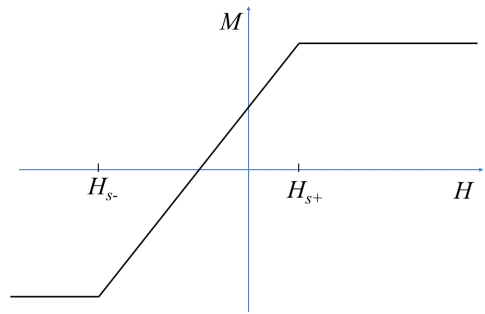}
\caption{\label{hard} Hysteresis loop of a biased hard-axis Stoner-Wohlfarth particle.}
\end{center}
%\label{figure:MI} this causes references to give section number, not figure number!!
%\vspace{-3mm}
\end{figure}

Our objective is to extract information separately for irreversible and reversible parts of a system.  Conceptually, it is simplest to think of an "easy-hard mixture" of Stoner-Wohlfarth particles with their easy and hard axes along the field, respectively\cite{NewellSW}.  The easy axis particles switch completely irreversibly at some fields $H_R$ and $H$ as in Fig. \ref{hyst}, and the hard axis particles switch reversibly: $M(H)$ is exactly linear until it saturates at some "saturation fields" $H_{s+}$ and $H_{s-}$ (Fig. \ref{hard}), which can have different magnitudes if we allow a bias.  

We have shown that the Preisach distribution completely describes the irreversible particles.  In a reversible system, on the other hand, if we change the magnetic moment by lowering the field from $H$ to $H_R$ and then raise it to $H_R$ again, this reverses the magnetization change and we return to the same magnetization at $H$, independently of $H_R$.  That is, the derivative with respect to $H_R$ (which we have denoted by $M_R$) is exactly zero, as is the second derivative -- the FORC distribution is exactly zero.

This makes it clear that the usual FORC distribution $\rho(H_R,H)$ does not completely determine the original FORC function $M(H_R,H)$.  To get a function by integrating its derivatives, one needs boundary conditions.  It turns out that we can do one integration: we can get the first derivative $M_R$ from the second derivative, by adding the plaquettes in the green  region of Fig. \ref{Strip}, because the boundary condition at the other end ($H \rightarrow \infty$) is known: $M=M_s$ in this limit, so all derivatives, including $M_R$, are zero.

\begin{figure}[htb]
\begin{center}
\includegraphics[width=3in]{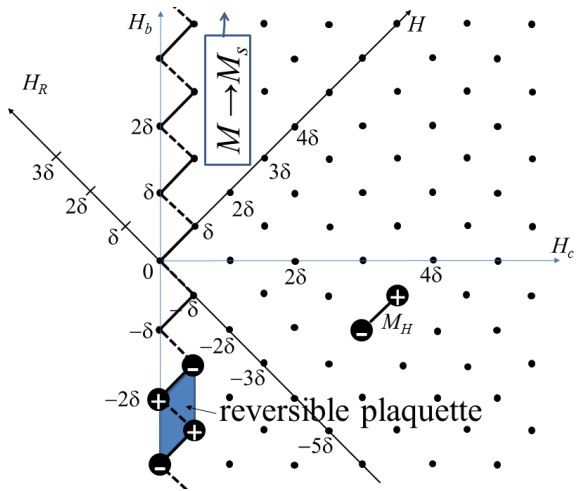}
\caption{\label{zig} Graphical demonstration that the derivative $M_H$ along the left boundary ($H=H_R+\frac{1}{2}\delta$), together with the other derivative $M_R$ which is determined by the irreducible distribution $\rho^{irr}(H_R,H)$, uniquely determines the FORC function along the zig-zag line at the left boundary, and therefore the entire FORC function $M(H_R,H)$. Also, $M_H$ can be determined everywhere by adding the moments of the "reversible plaquettes" along the left boundary,
% which are proportional to the reversible distribution $\rho^{rev}(H)$, 
since we know the boundary condition $M \rightarrow M_s$ at the top. }
\end{center}
%\label{figure:MI} this causes references to give section number, not figure number!!
%\vspace{-3mm}
\end{figure}

Knowing $M_R$, we could obtain $M$ everywhere by integrating along the $H_R$ axis, if we knew a boundary condition on $M$.  We do not know this at the lower right ($H_R \rightarrow -\infty$), but it would be sufficient to know it along the $H=H_R$ boundary (the $H_b$ axis).  But this is just the usual hysteresis loop, which contains both irreversible and reversible information -- we want to separate these.  However, the other first derivative, $M_H \equiv \partial M(H_R,H) / \partial H$ vanishes exactly along this boundary in an irreversible system (if the coercivity is at least $\delta$), so this is a candidate for describing the reversible part.  It contains all the rest of the information in the FORC function, in the sense that $M(H_R,H_R)$ along the boundary can be obtained by alternately adding $M_R$ and $M_H$ along a zig-zag path along the $H_b$ axis (Fig. \ref{zig}).

The treatment of irreversible effects has been discussed extensively in the literature -- in an "extended FORC" distribution\cite{Pike2003} this function is added to the irreversible FORC as a Dirac delta function at zero coercivity.  In this paper, however, we want to separate the reversible and irreversible behaviors.  In a model system consisting of a single easy axis (irreversible) and a single hard axis (reversible) particle we would like to get a single Dirac delta function in each of the irreversible and reversible FORC distributions, each giving the properties of the corresponding particle.  To this end, we note that a similar function has been used to extract anisotropy distributions of hard-axis systems\cite{d2mdh2,Lu}.  Since $M(H)$ is linear for a hard-axis particle (Fig. \ref{hard}), its derivative $dM/dH$ is a step function, and the second derivative has Dirac delta functions at $H=H_{s-}$ and $H=H_{s+}$.  Thus if there is a distribution of $H_{s-}$ and $H_{s+}$, the corresponding part of $d^2M/dH^2$ is proportional to this distribution.  More precisely,
$ - H d^2M(H,H)/dH^2 \approx \rho_{s+}(H) - \rho_{s-}(H)$.   However, this cannot be used for a mixed system because $d^2M/dH^2$ will be contaminated by the irreversible particles.  To obtain a distribution describing only reversible particles, we must start instead with 
\begin{equation}\label{MH}
M_H(H_R,H)\equiv \frac{1}{\delta} [ M(H_R, H+\frac{1}{2}\delta) - M(H_R, H-\frac{1}{2}\delta) ]
\end{equation}
which we have shown vanishes near the $H=H_R$ boundary for an irreversible system.
The signs of these two terms for the point $M_R=-4\delta, H=2.5\delta$ labeled "$M_H"$ in Fig. \ref{zig} are indicated by + and - signs.  
For a hard-axis reversible particle, $M(H_R,H)$ is independent of $H_R$ and linear in $H$, so $M_H(H_R,H_R+\frac{1}{2}\delta)$ is constant except at the saturation fields.  Thus the second derivative indicated by the "reversible plaquette" in Fig. \ref{zig} vanishes except at the saturation fields, and can be regarded as the saturation field distribution of the reversible particles:
\begin{multline}\label{rhorev}
\rho^{rev}(H) = \\
- \frac{H}{\delta ^2} [M(H,H) - M(H,H+\delta) + M(H+\delta,H) - M(H+\delta,H+\delta)]
\end{multline}
For $H<0$, this is negative and gives the distribution of $H_{s-}$; for $H>0$, it is positive and gives the distribution of $H_{s+}$. Note, however, that it does not give the joint distribution of $H_{s-}$ and $H_{s+}$, in the way that $\rho^{irr}$ gives the joint distribution of $H_R$ and $H$.  

\section{Visualization}
The most straightforward way to visualize the discrete irreversible FORC distribution is to paint each plaquette with a color density proportional to $\rho(H_R,H)$.  Even if the data is noisy, so plaquettes with high density are next to ones with low density, this scheme takes advantage of the natural averaging capability of the human eye: if one looks at such a display from a little further way, the fluctuations average out, in a way that they would not if we used a color-coding other than density (intensity)..  However, most commercial visualization software is not designed to display uniform-color plaquettes: it wants to interpolate the color continuously, which in the present case just obscures the simplicity of the discrete distribution.  For example, a very sharp peak will give density in only a single plaquette -- this sharpness will be obscured by interpolation or averaging.

The only way to directly control the color of each plaquette is to code the visualization at the lowest level -- currently all computer displays use OpenGL functions to display "primitives" (triangles, in our case).  Accordingly, we are working on a C++ code that uses direct calls to OpenGL functions \cite{FORC+}.  It is well known that commercial visualization software that is usually used to visualize the FORC distribution, which uses interpolation or extrapolation, can create artifacts, especially at the boundaries of the displayed region, which make it appear that there is a nonzero density when in fact it is almost zero.  This can occur along artificial boundaries, {\it i. e.}, at the ends of the FORC curves or along the last FORC curve (with largest or smallest $H_R$), or at the natural boundary $H_R=H$ ({\it i. e.}, $H_c=0$).  At the natural boundary, extrapolation can also lead to the mixing of irreversible and reversible effects, which our scheme separates cleanly.  In some cases, the data might be so noisy that the averaging capability of the eye is not enough -- then we can use a smoothing procedure (for example, fitting to a polynomial before extracting the mixed partial derivative)\cite{Egli}.

%\section{Future work}
%Useful extensions of the FORC+ program would be to include mean-field effects \cite{Zimanyi}.

\section{Conclusion}
In this paper we have derived a method for FORC analysis that optimally separates reversible and irreversible behavior -- it gives an irreversible FORC distribution $\rho^{irr}(H_R,H)$ that vanishes identically in a reversible system, and a reversible FORC distribution $\rho^{rev}(H_{sat})$ that vanishes identically in an irreversible system.  In a simple "easy-hard mixture" of Stoner-Wohlfarth particles, $\rho^{irr}$ completely describes the distribution irreversible (easy-axis) particles, and $\rho^{rev}$ completely describes the distributions of both upper and lower saturation fields.

%\section{Acknowledgements}
%

\end{document}